\documentclass[preprint]{ptephy_v2}

\preprintnumber{2509-04899} 
\usepackage{hyperref}

\usepackage{algorithm}
\usepackage{algorithmic}

\begin{document}

\title{Encoding of musical structures in hidden units of restricted Boltzmann machines}


\author{Mutsumi Kobayashi}

\author{Hiroshi Watanabe$^*$}
\affil{Department of Applied Physics and Physico-Informatics, Faculty of Science and Technology, Keio University, Yokohama 223-8522, Japan \email{hwatanabe@appi.keio.ac.jp}}




\begin{abstract}%
  Restricted Boltzmann machines (RBMs) are energy-based models originating from statistical physics, in which hidden units mediate the probability distribution of high-dimensional visible configurations. In this study, we use symbolic music as a structured non-physical dataset and investigate how musical regularities are encoded in the hidden layer of a Bernoulli-Bernoulli RBM. Musical scores by J.~S.~Bach are converted into binary piano-roll representations and used to train the model in an unsupervised manner. We then analyze the visible-layer patterns induced by individual hidden units by activating hidden units separately and computing the corresponding expected visible configurations. The trained RBM reconstructs piano-roll-like inputs and assigns lower visible free energies to the examined piano-roll configurations than to digit and noise inputs, establishing higher model probabilities for these particular piano-roll inputs without isolating which musical properties have been learned. The hidden units mainly encode local temporal and pitch-statistical structures, such as sparse piano-roll-like textures, rather than directly separable musical concepts such as melodies, chords, or keys. We also analyze hidden-layer representations using t-SNE and find that transposed versions of the same musical pieces are not necessarily mapped to nearby regions in the hidden space. Comparisons using input and hidden cosine similarities and PCA of the input piano rolls suggest that this organization reflects simple overlap between input patterns and sensitivity to absolute pitch, rather than demonstrating recognition of transposition-equivalent musical phrases. Samples from the trained RBM show local pitch organization, whereas iterative continuation reveals limited long-range coherence. These results provide a statistical-physics case study of how a simple spin model represents structured creative data and clarify both the usefulness and limitations of standard RBMs as interpretable models of musical structure.
\end{abstract}

\subjectindex{A22, A41, A42}
\maketitle

\section{Introduction}
\label{sec:introduction}

A central motivation for studying restricted Boltzmann machines (RBMs) is that the features learned by their hidden units can sometimes be interpreted in terms of recognizable parts of the visible data. In models trained on handwritten-digit images, Hinton reported that some hidden-unit receptive fields resemble localized fragments of pen strokes~\citep{Hinton2002}. More recently, Liao et al.\ also observed stroke-like filters in the learned weights of Gaussian--Bernoulli RBMs trained on MNIST~\citep{Liao2022}. A hidden unit can thus be associated with a visible pattern that is meaningful to a human observer. These observations motivate our investigation of the musical counterparts of these strokes in an RBM trained on musical scores.

RBMs provide a direct connection between this question and music modeling. Boulanger-Lewandowski et al.\ used RBMs to model correlated patterns of simultaneous notes and combined them with recurrent networks to generate polyphonic music~\citep{Boulanger2012}. Their examples of visible conditional expectations contain recognizable harmonic configurations. These results suggest that RBM parameters can encode musically meaningful regularities. They motivate the expectation that an RBM trained on music might acquire hidden units associated with chord structures or characteristic short phrases, just as hidden units in a handwritten-digit model can acquire stroke-like features. Whether such musical structures can actually be identified in individual hidden units is the issue we investigate.

Our purpose is to examine what musical information is stored in an RBM that has been trained to reconstruct and generate piano-roll configurations. In a piano roll, simultaneous notes form patterns along the pitch axis, while short melodic phrases form patterns extending across pitch and time. We therefore ask whether individual hidden units store visible pitch--time structures that a human observer can recognize as components of music. Here, ``stored'' refers to patterns encoded in the learned weights and expressed through the conditional expectation of the visible layer when a hidden configuration is specified; it does not imply that a complete piece is assigned to a single unit. Chords and short phrases are candidate structures motivated by the stroke analogy, whose presence must be examined rather than assumed.

We use a standard Bernoulli--Bernoulli RBM trained on two-measure piano-roll images of keyboard works by J.~S.~Bach. An RBM is an energy-based model with visible and hidden layers and no connections within either layer~\citep{Smolensky1986,Ackley1985,Zhang2018}. This structure makes the association between each hidden unit and the visible layer directly accessible through its weights and conditional expectations. Representing two measures within a single visible vector allows both simultaneous pitch configurations and short temporal patterns to be examined in the same model. The resulting analysis asks how the learned representation relates to musical structures at the scale of this training window.

Related work has examined musical organization in the hidden space. Cancino Chac\'on et al.\ applied principal component analysis to the hidden states of an RBM trained on musical $n$-grams and found tonal relationships associated with the circle of fifths~\citep{Chacon2014}. We focus on the visible patterns associated with individual hidden units in a model trained directly on piano rolls.

The analogy with image features also raises a question about translation. In a standard fully connected RBM, the weights are tied to absolute pixel positions, so a feature and its translated copy can produce different hidden responses. The architecture does not impose translation invariance. Lee et al.\ addressed this issue in image modeling by introducing convolutional deep belief networks with shared feature detectors and probabilistic max-pooling~\citep{Lee2009}. Their approach illustrates the importance of representing a feature across different input locations.

In a piano-roll image, a translation along the pitch axis corresponds to musical transposition. Shifting all notes by the same number of semitones moves the pattern vertically while preserving its timing and relative pitch intervals, provided that the notes remain within the represented pitch range. A chord or short phrase can therefore retain its musical relationships while occupying different image pixels. A fully connected RBM may nevertheless respond differently because its weights refer to absolute pitches~\citep{Johnson2017}. We therefore investigate how the hidden representation of the trained RBM responds to pitch-axis translations of the same piano-roll input. In particular, we ask whether the representation retains similarities between transposed versions, while distinguishing such relationships from strict invariance of the hidden state.

Specifically, we address the following three questions.
\begin{enumerate}
  \item What pitch--time patterns are associated with individual hidden units, and do these patterns resemble chords or characteristic short phrases, as stroke-like features do in handwritten-digit models?
  \item How do hidden representations change under pitch transposition, and what does this reveal about the representation of the same musical pattern at different pitches?
  \item What local musical organization is visible in reconstructed and generated piano rolls, and what limitations become apparent when generation is continued beyond the training window?
\end{enumerate}

The experiments are organized around these questions. Reconstruction and energy comparisons with musical and non-musical inputs provide preliminary checks of the trained model's responses; they do not by themselves identify which musical structures are represented. The main analysis activates hidden units individually and visualizes the corresponding visible expectations to look for musical counterparts of strokes. The transposition experiment uses t-distributed stochastic neighbor embedding (t-SNE) to explore changes in the joint hidden representation. Finally, generated piano rolls and iterative continuation show the visible configurations produced by the learned model under the specified generation rule. These complementary observations connect the hidden-unit analysis to the model's observable behavior.

The contribution of this study is a direct examination of the relationship between hidden-unit patterns and human-recognizable musical components in a piano-roll RBM. The visualized patterns contain local pitch--time regularities, while recognizable chords and phrases are not readily apparent in the individual-unit examples. Such observations concern the interpretability of the examined units and do not exclude the encoding of musical information through combinations of units. Our aim is to characterize what is visible in this trained model's representation and where the analogy with stroke-like parts is limited. The generation experiments serve this analysis of learned information, rather than a comparison of composition quality with more elaborate music models.

The remainder of this paper is organized as follows. Section~\ref{sec:method} describes the RBM, the piano-roll dataset, and the analysis procedures. Section~\ref{sec:results} presents the reconstruction and energy analyses, hidden-unit visualizations, transposition analysis, and generated configurations. Section~\ref{sec:summary} summarizes the results and discusses their implications.

\section{Methods}\label{sec:method}

\subsection{RBM}

In this study, we use an RBM as a minimal energy-based model for analyzing structured binary data. An RBM is a restricted form of the Boltzmann machine, a stochastic spin model whose probability distribution is determined by an energy function~\citep{Hinton1983, Ackley1985}. The restriction is that units within the same layer are not directly connected. Interactions exist only between the visible layer, which represents observed data, and the hidden layer, which mediates latent statistical features. This bipartite structure makes the model simpler than a general Boltzmann machine and allows efficient training by alternating conditional sampling between the two layers. In this work, we adopt a Bernoulli-Bernoulli RBM, in which both visible and hidden units are binary variables taking values of either $0$ or $1$~\citep{Yamashita2014}.

We consider an RBM model with $D$ visible units and $P$ hidden units. The states of the visible and hidden layers are denoted by $\boldsymbol{v} = \{v_1, v_2, \cdots, v_D\}$ and $\boldsymbol{h} = \{h_1, h_2, \cdots, h_P\}$, respectively, where each $v_i$ and $h_j$ takes a binary value of either $0$ or $1$. The energy of the model with the states $\{\boldsymbol{v}, \boldsymbol{h}\}$ is given by,
\begin{equation}
  E(\boldsymbol{v}, \boldsymbol{h}|\theta) = - \sum_{i=1}^{D} \sum_{j=1}^{P} w_{ij} v_i  h_j - \sum_{i=1}^{D} v_i b_i - \sum_{j=1}^{P} h_j c_j, \label{eq:energy_function}
\end{equation}
where $\theta = \{w_{ij}, b_i, c_i\}$ are the model parameters~\citep{Zhang2018}. The weights $\boldsymbol{W} = \{w_{ij}\}$ represent the interactions between the visible unit $v_i$ and the hidden unit $h_j$. The parameters $\boldsymbol{b} = \{b_i\}$ and $\boldsymbol{c} = \{c_i\}$ represent the biases of the visible and hidden layers, respectively. Given the model parameter $\theta$, the probability of the visible layer being in the state $\boldsymbol{v}$ is given by,
\begin{equation}
  p(\boldsymbol{v}|\theta) = \frac{\sum_{\boldsymbol{h}}\exp\{-E(\boldsymbol{v},\boldsymbol{h}|\theta)\}}{Z(\theta)}, \label{eq:model_distribution}
\end{equation}
where $Z(\theta)$ is the partition function defined by
\begin{equation}
  Z(\theta) = \sum_{\boldsymbol{v}} \sum_{\boldsymbol{h}} \exp\{-E(\boldsymbol{v},\boldsymbol{h}|\theta)\}.
\end{equation}

For a fixed visible input, the visible free energy is defined by summing over all hidden states,
\begin{equation}
  \begin{aligned}
    F(\boldsymbol{v}|\theta)
     & = -\log\sum_{\boldsymbol{h}}\exp\{-E(\boldsymbol{v},\boldsymbol{h}|\theta)\} \\
     & = -\sum_{i=1}^{D} b_i v_i
    -\sum_{j=1}^{P}\log\left[1+\exp\left(c_j+\sum_{i=1}^{D}w_{ij}v_i\right)\right].
  \end{aligned}
  \label{eq:visible_free_energy}
\end{equation}
Thus, \(\log p(\boldsymbol{v}|\theta)=-F(\boldsymbol{v}|\theta)-\log Z(\theta)\). Under the same trained model, a lower visible free energy corresponds to a higher input probability, and differences in log-likelihood can be obtained from free-energy differences without computing the partition function. This quantity differs from the joint energy averaged over hidden states sampled conditionally on a fixed visible input, \(\langle E(\boldsymbol{v},\boldsymbol{h}|\theta)\rangle_{\boldsymbol{h}\mid\boldsymbol{v}}\).

The bipartite structure of the RBM also gives a direct way to interpret individual hidden units. If one hidden unit $h_j$ is fixed to $1$ and all other hidden units are fixed to $0$, the conditional expectation value of the visible unit $v_i$ is
\begin{equation}
  \langle v_i \rangle_{h_j=1, h_{k \neq j}=0}
  = \sigma(b_i + w_{ij}),
  \label{eq:hidden_spin_visible_expectation}
\end{equation}
where $\sigma(x) = 1/(1+\exp(-x))$ is the sigmoid function. Thus, the weight vector connected to a hidden unit can be mapped back to the visible layer as a visible pattern. In the present study, we use this property to examine what kind of piano-roll structures are encoded by individual hidden units after training.

The goal of training an RBM is to make the model distribution $p(\boldsymbol{v}|\theta)$ approximate the data distribution $q(\boldsymbol{v})$ as closely as possible. To achieve this goal, we adopt the Kullback-Leibler (KL) divergence as the cost function. The KL divergence between the model distribution $p(\boldsymbol{v}|\theta)$ and the data distribution $q(\boldsymbol{v})$ is defined by
\begin{equation}
  \mathrm{KL}\left[ q(\boldsymbol{v}) \boldsymbol{|} p(\boldsymbol{v}|\theta) \right]
  = \sum_{\boldsymbol{v}} q(\boldsymbol{v}) \log \frac{q(\boldsymbol{v})}{p(\boldsymbol{v}|\theta)}.
\end{equation}
The exact model expectation in the gradient of this objective is generally intractable. We approximate the update using one-step Contrastive Divergence (CD-1)~\citep{Hinton2002}, rather than evaluating the full-distribution KL divergence directly. We implemented the RBM model using Python, and by utilizing CuPy, we were able to accelerate the computations through GPGPU processing~\citep{Nishino2017}. The RBM code developed in this study is available on GitHub~\citep{code}.

\subsection{Dataset}

\begin{figure}[htbp]
  \centering
  \includegraphics[width=15cm]{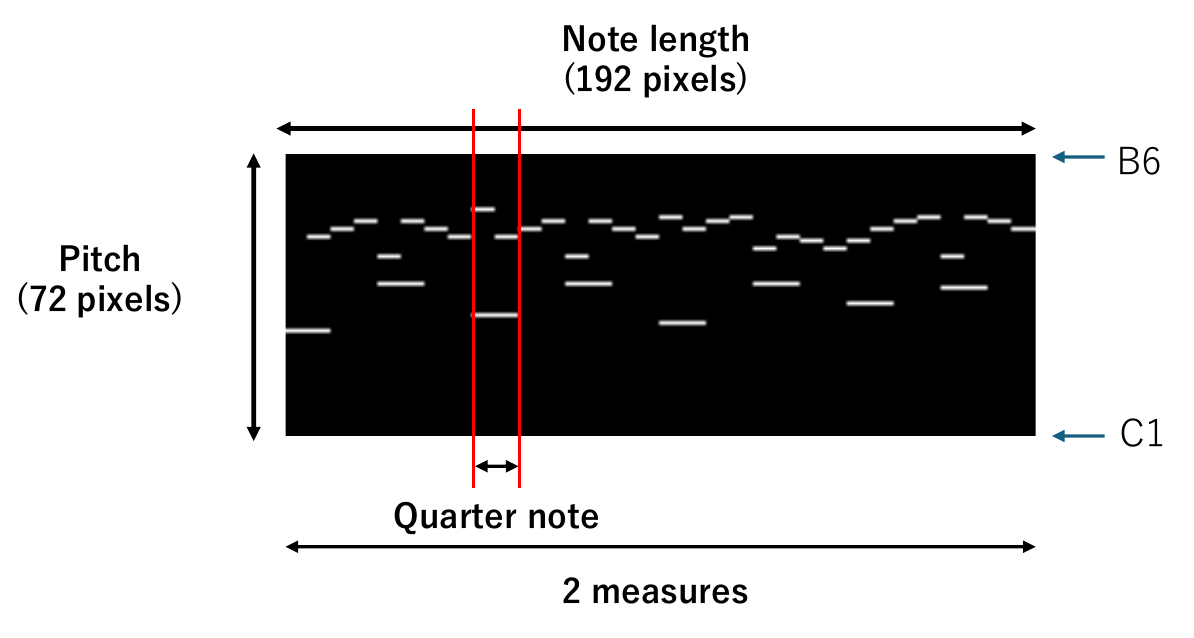}
  \caption{(Color online) Piano roll representation of a musical segment. The horizontal axis represents time (note duration), with a total width of 192 pixels corresponding to two measures in 4/4 time. The vertical axis represents pitch, spanning 72 pixels from C1 (the lowest C on a standard 88-key piano) to B6. Each horizontal bar indicates a note, with its vertical position corresponding to pitch and its horizontal length indicating duration. A quarter note is represented by 24 pixels in width.}
  \label{fig:pianoroll}
\end{figure}

As a structured binary dataset, we used piano-roll representations of compositions by J.~S.~Bach. A total of 56 MIDI files were obtained from the Mutopia Project~\citep{mutopia}, and each file was converted into a black-and-white piano-roll image. The source identifiers and licensing information are recorded in the dataset manifest distributed with the code. A piano roll is a two-dimensional representation of musical information, where the horizontal axis corresponds to time and the vertical axis corresponds to pitch. Notes are depicted as horizontal bars, with their positions and lengths indicating the timing and duration of each note, respectively. Each pixel value in the piano-roll image is either 0 or 1 and is directly assigned to a visible unit of the Bernoulli-type RBM. In this representation, a musical segment is treated as a binary visible configuration with structured correlations along both the time and pitch directions.

To standardize the input dimensions, we used a four-quarter-note measure and extracted two-measure windows with a stride of one measure. The final window was padded with zeros when fewer than two measures remained. MIDI tempos were normalized to 120 beats per minute, and sustain-pedal events were ignored when constructing the binary piano rolls.

The image size was fixed at $72 \times 192$ pixels. The vertical dimension of 72 pixels corresponds to the pitch range from C1 to B6, where C1 denotes the C note in the first octave of a standard 88-key piano (i.e., the lowest C key), and B6 denotes the B note in the sixth octave, one semitone below the highest C (C8). The horizontal dimension of 192 pixels represents time, with 24 pixels corresponding to the duration of one quarter note. This resolution was chosen so that the horizontal pixel count would be divisible by 3, enabling the representation of triplet notes. Image size and the number of hidden units used for training are summarized in Table~\ref{tab:settings}.

Under this specification, the shortest representable note value is a sixty-fourth-note triplet. While this prevents accurate representation of regular sixty-fourth notes—which are the shortest notes found in the training data—their occurrence in the dataset was negligible. Moreover, this specification was adopted to keep the size of the training data computationally manageable.

Each window was represented at 12 pitch offsets, comprising the original and transpositions from five semitones below to six semitones above it. Transposition was applied after cropping to the represented pitch range; shifted pixels outside this range were discarded, without wraparound. This preprocessing produced 21,444 binary training images. All images were used for training, without a held-out validation split.

\begin{table}[htbp]
  \centering
  \caption{Model dimensions and training conditions}
  \label{tab:settings}
  \begin{tabular}{lc}
    \hline
    Image size             & $72 \times 192$ \\
    Number of visible units & 13,824         \\
    Number of hidden units & 2048            \\
    Training images        & 21,444          \\
    Training method        & CD-1            \\
    Training epochs        & 1,000           \\
    Configured batch size  & 500             \\
    Learning rate          & $10^{-4}$       \\
    Momentum               & 0.95            \\
    Interlayer connections & Fully connected \\
    \hline
  \end{tabular}
\end{table}

\subsection{Training procedure and progress diagnostic}

The RBM has 13,824 visible units and 2,048 hidden units and was trained for 1,000 epochs with the conditions in Table~\ref{tab:settings}. We initialized weights uniformly in $[-\sqrt{6/(D+P)},\sqrt{6/(D+P)}]$, with zero biases and momentum increments. All visible--hidden connections were retained, without connection masking. We used a fixed training random seed and shuffled the training images at each epoch. The configured batch size was 500. The implementation divides the shuffled dataset into $\lfloor 21444/500\rfloor=42$ nearly equal batches, so the actual batch sizes were 510 or 511, with all images included in each epoch. Training was performed with CuPy on an NVIDIA A100 GPU.

For each batch, we sampled $\boldsymbol{h}_0$ conditionally on the input $\boldsymbol{v}_0$, sampled $\boldsymbol{v}_1$ conditionally on $\boldsymbol{h}_0$, and sampled $\boldsymbol{h}_1$ conditionally on $\boldsymbol{v}_1$. The weight update used the batch sum of $\boldsymbol{v}_0\boldsymbol{h}_0^\mathsf{T}-\boldsymbol{v}_1\boldsymbol{h}_1^\mathsf{T}$, while the bias updates used batch means of the corresponding state differences. Following the reference implementation, the parameter increment was $\Delta\theta\leftarrow\mu\Delta\theta+(1-\mu)\eta g_\theta$, followed by $\theta\leftarrow\theta+\Delta\theta$, with learning rate $\eta=10^{-4}$ and momentum $\mu=0.95$.

We monitored a reconstruction-based diagnostic rather than the exact KL divergence between the data and model distributions. For each input bit, we evaluated the Bernoulli KL divergence between its binary value and the reconstructed activation probability $r_{ni}=p(v_i=1\mid\boldsymbol{h}_{0,n},\theta)$, using the sampled hidden state and the parameters after the batch update. For a batch of size $B$ and $D=13,824$ visible units, the reported value was
\begin{equation}
  L_B=-\frac{1}{BD}\sum_{n=1}^{B}\sum_{i=1}^{D}
  \left[v_{ni}\log r_{ni}+(1-v_{ni})\log(1-r_{ni})\right]
  -\frac{\log B}{D}.
  \label{eq:training_diagnostic}
\end{equation}
For numerical stability, the probabilities in this diagnostic were clipped to $[10^{-7},1-10^{-7}]$. The final term reproduces the batch-size normalization correction in the reference implementation. The epoch diagnostic is the mean of $L_B$ over the 42 batches. It decreased from approximately $0.264$ at the first epoch to $8.86\times10^{-4}$ at epoch 1,000. Although it continued to decrease slowly at late epochs, its substantial reduction in absolute magnitude was judged sufficient for the present analyses. This is an operational assessment of reconstruction progress and does not establish convergence of the model likelihood or the full model distribution.

\subsection{Sampling from the trained RBM}

Having trained a simple RBM on piano-roll images, we next investigate whether the learned model can be used to compose music. A direct approach is ordinary block Gibbs sampling, which alternates conditional Bernoulli draws of the hidden and visible units. However, for the trained model and the initial conditions examined here, this procedure did not produce coherent musical configurations. Random initialization yielded noise-like images, while all-zero initialization led to only sparse horizontal traces. The corresponding results are presented in Appendix~\ref{sec:standard_gibbs}. We therefore investigate whether modifying the generation algorithm can produce piano rolls with recognizable musical structure from the same trained RBM.

\begin{figure}[htbp]
  \centering
  \includegraphics[width=\linewidth]{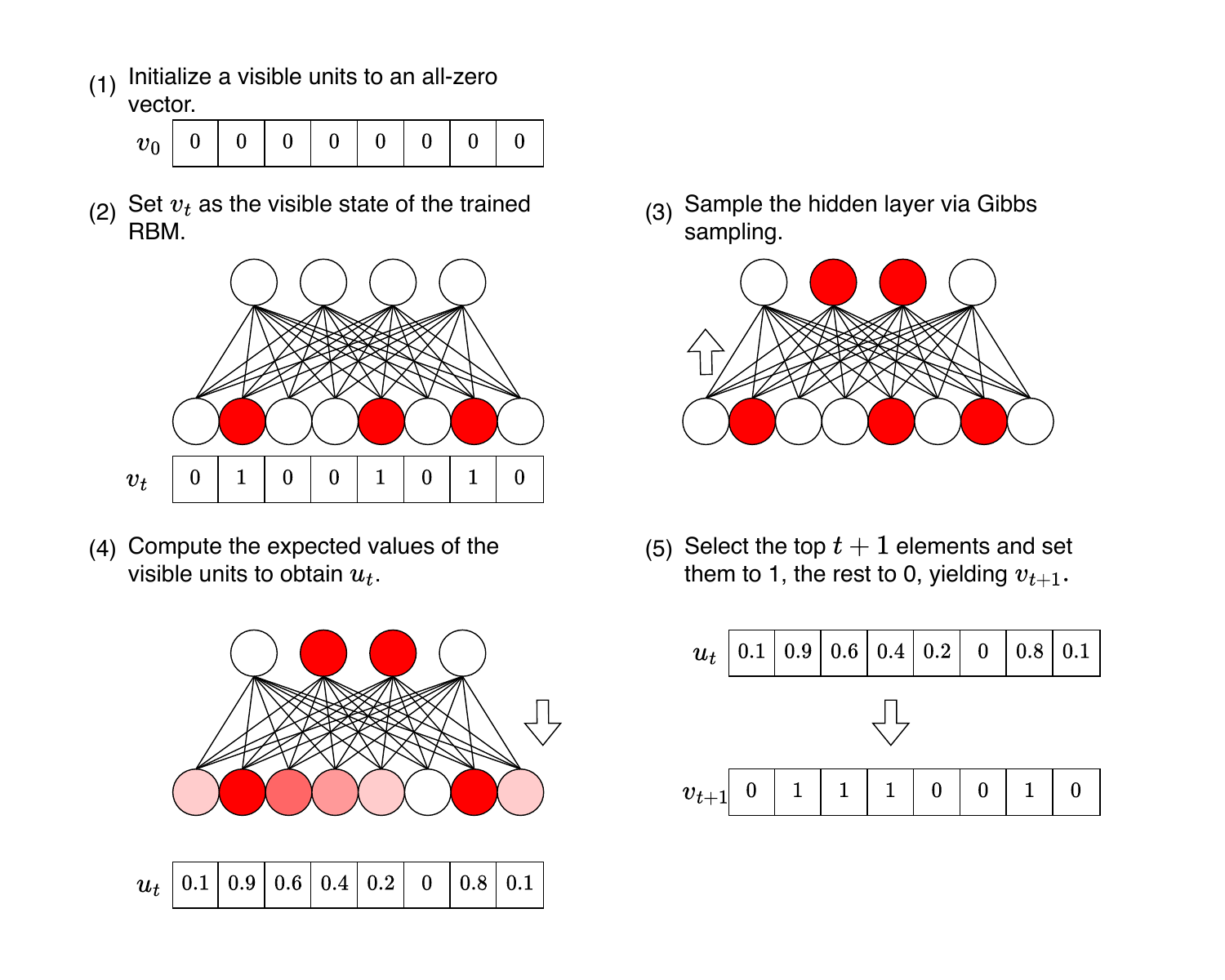}
  \caption{(Color online) Schematic illustration of the activity-controlled piano-roll generation procedure using the trained RBM. Step (1) is performed only once at initialization, whereas steps (2)--(5) form a repeated loop. The illustrated iteration is \(t=3\), producing a visible configuration with four active units from one with three active units.}
  \label{fig:algorithm}
\end{figure}

Our modified procedure retains conditional sampling of the hidden units but replaces the Bernoulli draw of the visible units with a selection based on their conditional expectations. At each generation step, the visible units with the largest expectations are activated, and the prescribed number of active units is increased by one. This allows us to construct increasingly populated piano rolls using the patterns encoded in the learned model. The aim is to demonstrate that a suitable generation rule can draw out local musical organization that is not apparent in the ordinary Gibbs samples shown in Appendix~\ref{sec:standard_gibbs}. The procedure is illustrated in Fig.~\ref{fig:algorithm} and detailed in Algorithm~\ref{proc:algorithm1}. Because the visible update is externally constrained, the generated configurations are not ordinary Gibbs samples from the RBM distribution.

\begin{algorithm}
  \caption{Activity-Controlled Generation of a Two-Measure Piano Roll}
  \label{proc:algorithm1}
  \begin{algorithmic}[1]
    \STATE \textbf{Input:} Trained RBM parameters \(\theta\), visible dimension \(D\), and target number \(N\) of active visible units.
    \STATE \textbf{Output:} Binary visible vector \(\mathbf{v}_N \in \{0,1\}^{D}\) with \(N\) active units.
    \STATE Initialize the visible state as \(\mathbf{v}_0 = \mathbf{0}\).
    \FOR{\(t = 0, 1, \ldots, N-1\)}
    \STATE Sample the hidden state \(\mathbf{h}_t\) from \(p(\mathbf{h}|\mathbf{v}_t,\theta)\).
    \STATE Compute the conditional expectation \(\mathbf{u}_t = \langle \mathbf{v} \rangle_{\mathbf{h}_t}\).
    \STATE Construct \(\mathbf{v}_{t+1}\) by setting the \(t+1\) largest elements of \(\mathbf{u}_t\) to \(1\) and all remaining elements to \(0\).
    \ENDFOR
    \STATE Return \(\mathbf{v}_N\).
  \end{algorithmic}
\end{algorithm}

\begin{figure}[htbp]
  \centering
  \includegraphics[width=\linewidth]{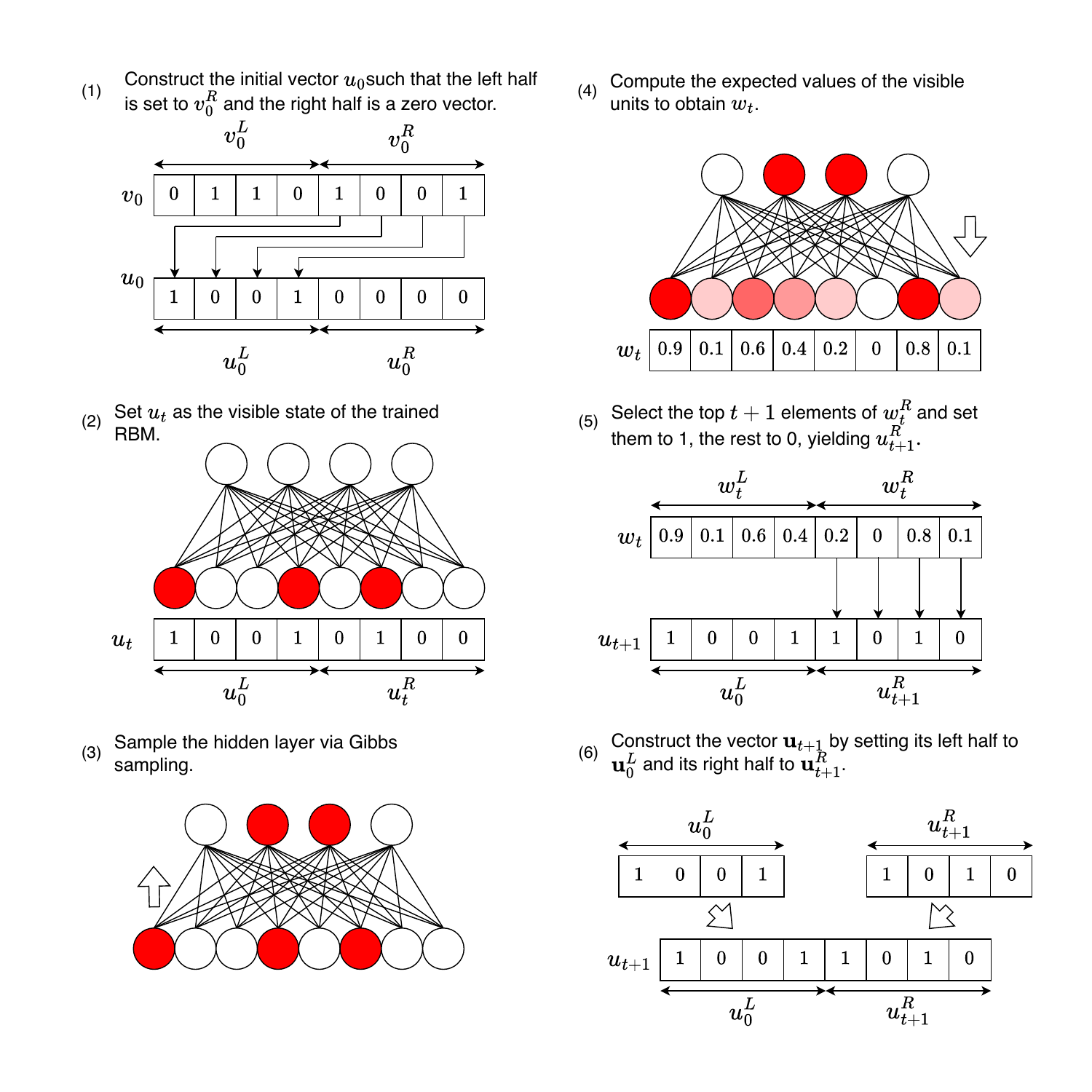}
  \caption{(Color online) Schematic illustration of the continuation procedure for an already generated piano-roll configuration.}
  \label{fig:algorithm2}
\end{figure}

The number of visible units in the RBM used for training was \(13,824\), which corresponds to two measures in 4/4 time. Algorithm~\ref{proc:algorithm1} therefore generates a two-measure piano-roll configuration. To extend the generated music beyond this window, we also use an iterative continuation procedure. The latter measure of a generated two-measure piano roll is fixed as the first measure of the next window, and the remaining measure is generated using the same activity-controlled principle. This extension procedure is illustrated in Fig.~\ref{fig:algorithm2} and detailed in Algorithm~\ref{proc:algorithm2}.

\begin{algorithm}
  \caption{Continuation of a Generated Piano Roll}
  \label{proc:algorithm2}
  \begin{algorithmic}[1]
    \STATE \textbf{Input:} Trained RBM parameters \(\theta\), an initial two-measure vector \(\mathbf{v}_0 \in \{0,1\}^{13824}\), maximum activity \(N\) explored in each new measure, and number of extensions \(K\).
    \STATE \textbf{Output:} Piano-roll sequence \(\mathcal{S}\) containing \(K+2\) measures.
    \STATE Initialize \(\mathbf{v} = \mathbf{v}_0\) and \(\mathcal{S}\) with the two measures of \(\mathbf{v}_0\).
    \FOR{\(k = 1, 2, \ldots, K\)}
    \STATE Split the current window into one-measure halves \(\mathbf{v} = [\mathbf{v}^L, \mathbf{v}^R]^T\), each containing 6912 units.
    \STATE Initialize \(\mathbf{u}_0^L = \mathbf{v}^R\), \(\mathbf{u}_0^R = \mathbf{0}\), and \(\mathbf{u}_0 = [\mathbf{u}_0^L, \mathbf{u}_0^R]^T\).
    \STATE Initialize \(E_{\min}=+\infty\) and \(\mathbf{u}_{\mathrm{best}}=\mathbf{u}_0\).
    \FOR{\(t = 0, 1, \ldots, N-1\)}
    \STATE Sample the hidden state \(\mathbf{h}_t\) from \(p(\mathbf{h}|\mathbf{u}_t,\theta)\).
    \STATE Compute the conditional expectation \(\mathbf{w}_t = \langle \mathbf{v} \rangle_{\mathbf{h}_t}\).
    \STATE Split \(\mathbf{w}_t\) into one-measure halves as \(\mathbf{w}_t = [\mathbf{w}_t^L, \mathbf{w}_t^R]^T\).
    \STATE Construct \(\mathbf{u}_{t+1}^R\) by setting the \(t+1\) largest elements of \(\mathbf{w}_t^R\) to \(1\) and all remaining elements to \(0\).
    \STATE Set \(\mathbf{u}_{t+1} = [\mathbf{u}_0^L, \mathbf{u}_{t+1}^R]^T\), keeping the left half fixed.
    \STATE Evaluate \(E_{t+1}=E(\mathbf{u}_{t+1},\mathbf{h}_t|\theta)\).
    \IF{\(E_{t+1}<E_{\min}\)}
    \STATE Set \(E_{\min}=E_{t+1}\) and \(\mathbf{u}_{\mathrm{best}}=\mathbf{u}_{t+1}\).
    \ENDIF
    \ENDFOR
    \STATE Append only the newly generated measure \(\mathbf{u}_{\mathrm{best}}^R\) to \(\mathcal{S}\).
    \STATE Set \(\mathbf{v} = \mathbf{u}_{\mathrm{best}}\) as the current window for the next extension.
    \ENDFOR
    \STATE Return \(\mathcal{S}\).
  \end{algorithmic}
\end{algorithm}

Algorithm~\ref{proc:algorithm2} distinguishes two loops. The outer loop over \(k\) extends the sequence by one measure at a time. At the start of each extension, the right measure of the current two-measure window is copied to the fixed left half, \(\mathbf{u}_0^L = \mathbf{v}^R\), and the new right half is reset to zero. The inner loop over \(t\) generates that new measure by increasing its activity from zero to \(N\), while keeping the left half fixed. After exploring the full inner-loop trajectory, we select the state with the lowest joint energy \(E(\mathbf{u}_{t+1},\mathbf{h}_t|\theta)\), where \(\mathbf{h}_t\) is the sampled hidden state that generated \(\mathbf{u}_{t+1}\). Only the new right measure of this selected state is appended to the sequence, and the selected two-measure state becomes the current window for the next outer iteration. The random-number sequence is not rewound when an earlier state is selected. We use this continuation procedure to investigate whether the local musical organization of the generated piano rolls can be maintained in longer compositions.
The loss of musical coherence during repeated continuation is likely to depend on this generation algorithm, including its activity constraint and repeated fixing of the preceding measure, rather than solely on the properties of the trained RBM.
For the initial two measures, we run Algorithm~\ref{proc:algorithm1} up to \(N=1000\), retain the intermediate configurations, and select the state with the lowest joint energy along the trajectory. In Algorithm~\ref{proc:algorithm2}, we explore up to \(N=500\) active units in each new measure and use \(K=6\) outer iterations, producing an eight-measure piano roll without duplicating the overlapping measures. These values are exploration limits; the activity of the selected minimum-energy state can be lower and can vary between measures.

\section{Results}\label{sec:results}

\subsection{Reconstruction and Energy Evaluation}

\begin{figure}[htbp]
  \centering
  \begin{tabular}{@{}l@{\hspace{0.02\linewidth}}l@{}}
    \includegraphics[width=0.49\linewidth]{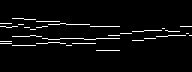} &
    \includegraphics[width=0.49\linewidth]{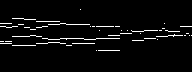} \\
    \includegraphics[width=0.49\linewidth]{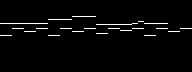} &
    \includegraphics[width=0.49\linewidth]{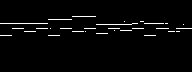}
  \end{tabular}
  \caption{Outputs obtained by Gibbs reconstruction from piano-roll inputs using the trained RBM.
    (a) Piano roll image of a J.~S.~Bach composition used for training.
    (a') Output obtained from (a) by Gibbs reconstruction.
    (b) Piano roll image of a W.~A.~Mozart composition not used during training.
    (b') Output obtained from (b) by Gibbs reconstruction.}
  \label{fig:reconstruct_piano_roll}
\end{figure}

To verify whether the trained RBM correctly memorized the piano rolls,
we input the piano roll into the visible units and examined whether it could be reconstructed through Gibbs sampling. For each input, the visible state is initialized with the binary input image, \(\mathbf{v}_0\). We sample \(\mathbf{h}_0 \sim p(\mathbf{h}\mid\mathbf{v}_0,\theta)\), then sample \(\mathbf{v}_1 \sim p(\mathbf{v}\mid\mathbf{h}_0,\theta)\), and display \(\mathbf{v}_1\) as the reconstruction. Thus, reconstruction uses exactly one block Gibbs step with the trained parameters held fixed. Both layers are sampled from their conditional Bernoulli distributions; the output is not obtained by thresholding conditional expectations, and no constraint is imposed on the number of active visible units. Figure \ref{fig:reconstruct_piano_roll} shows the input piano roll images and the images obtained through reconstruction.
First, when a piano roll of a J.~S.~Bach composition used during training was provided as input (Fig.~\ref{fig:reconstruct_piano_roll}~(a)), the RBM reconstructed a similar piano-roll configuration (Fig.~\ref{fig:reconstruct_piano_roll}~(a')). We also provided a piano roll of a W.~A.~Mozart composition that was not included in the training data (Fig.~\ref{fig:reconstruct_piano_roll}(b)). In this case, the reconstructed image also retained the overall piano-roll structure (Fig.~\ref{fig:reconstruct_piano_roll}(b')). These results show that, within this reconstruction setting, the trained RBM maps piano-roll-like inputs back to piano-roll-like visible configurations. In contrast, when digit images from the MNIST dataset were used as non-piano-roll inputs, the reconstructed outputs were noise-like rather than digit-like, as shown in Appendix~\ref{sec:mnist_reconstruction}. Thus, the trained RBM reconstructs inputs that match the piano-roll structure learned from the training data, while it does not reconstruct inputs outside this learned data structure.

To compare the trained model's responses to musical and non-musical inputs, we evaluated both the conditional mean joint energy and the visible free energy in Table~\ref{tbl:energy}. We used the fixed binary input images shown in Fig.~\ref{fig:reconstruct_piano_roll} and Appendix~\ref{sec:mnist_reconstruction}, representing Bach, Mozart, and MNIST digits 0, 5, and 8, together with one fixed noise image whose pixels were independently activated with probability \(1/2\). All inputs were represented as 0/1 vectors. For each fixed input \(\boldsymbol{v}\), we independently sampled ten hidden states from \(p(\boldsymbol{h}\mid\boldsymbol{v},\theta)\) and computed the mean and sample standard deviation of \(E(\boldsymbol{v},\boldsymbol{h}|\theta)\). The visible free energy was evaluated directly using Eq.~\eqref{eq:visible_free_energy}, without sampling or reconstructing the input.

Both piano-roll inputs have substantially lower mean joint energies and visible free energies than the digit and noise inputs examined. The free-energy comparison therefore establishes higher probabilities for these particular piano-roll configurations under the same trained model. Unlike the sampled joint-energy means, the visible free energies have no hidden-sampling uncertainty. For the fixed noise image, all ten joint-energy samples happened to be identical at the reported precision; the resulting zero sample standard deviation does not describe variability across different noise images. These comparisons are limited to the displayed inputs and do not by themselves distinguish learned musical relationships from simpler properties such as sparsity and horizontal continuity.

\begin{table}[htbp]
  \centering
  \caption{Conditional mean joint energy and visible free energy for fixed input images. Joint energies are reported as mean $\pm$ sample standard deviation (ddof=1) over ten independent hidden samples per fixed input. Visible free energies are evaluated analytically and have no hidden-sampling uncertainty. Noise denotes one fixed random binary image.}
  \begin{tabular}{lrrr}
    \hline
    Input                  & Active pixels & Joint energy         & Visible free energy \\
    \hline
    Piano roll (trained)   & 564           & $-3687.31 \pm 9.74$  & $-3785.65$          \\
    Piano roll (untrained) & 360           & $-3370.07 \pm 10.94$ & $-3590.73$          \\
    MNIST digit 0          & 2581          & $41.34 \pm 2.54$     & $38.72$             \\
    MNIST digit 5          & 1436          & $-332.61 \pm 7.12$   & $-368.88$           \\
    MNIST digit 8          & 1956          & $-33.42 \pm 2.21$    & $-43.27$            \\
    Noise                  & 7014          & $85.60 \pm 0.00$     & $85.60$             \\
    \hline
  \end{tabular}
  \label{tbl:energy}
\end{table}

\subsection{Visible-layer Patterns Induced by Individual Hidden Units}

\begin{figure}[htpb]
  \centering
  \makebox[0.05\linewidth]{(a)}\includegraphics[width=0.78\linewidth]{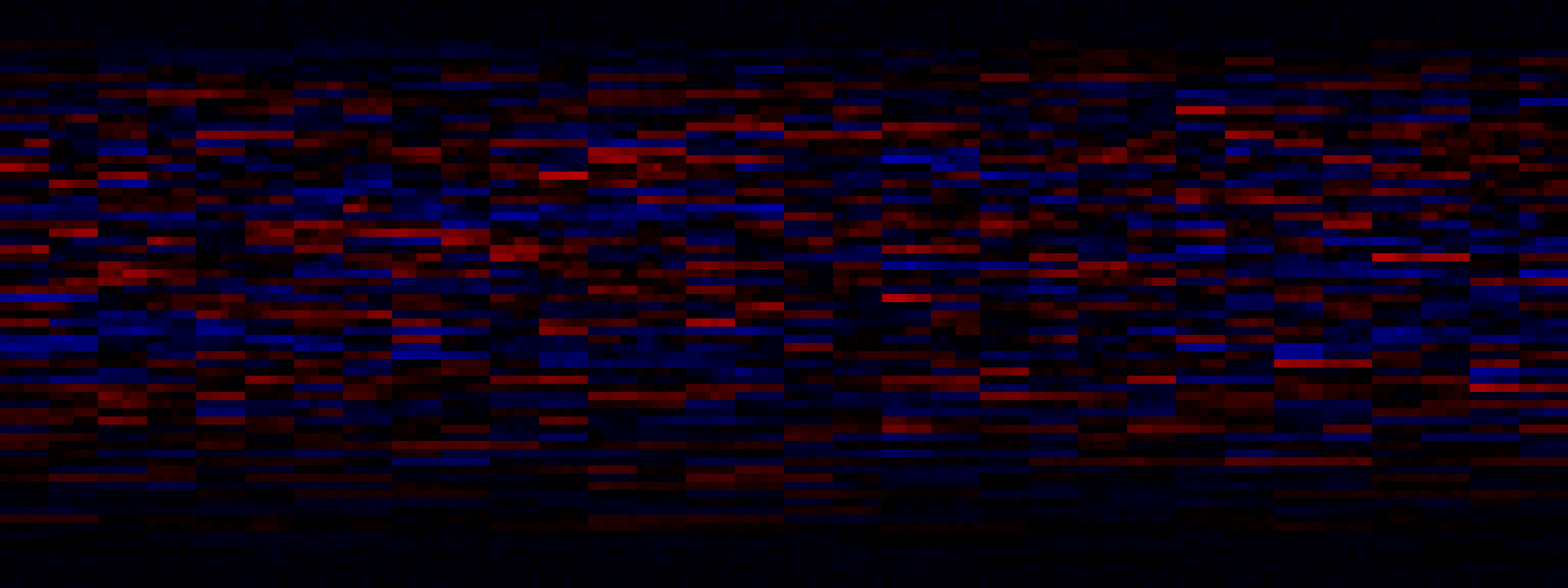}\par
  \makebox[0.05\linewidth]{}\includegraphics[width=0.78\linewidth]{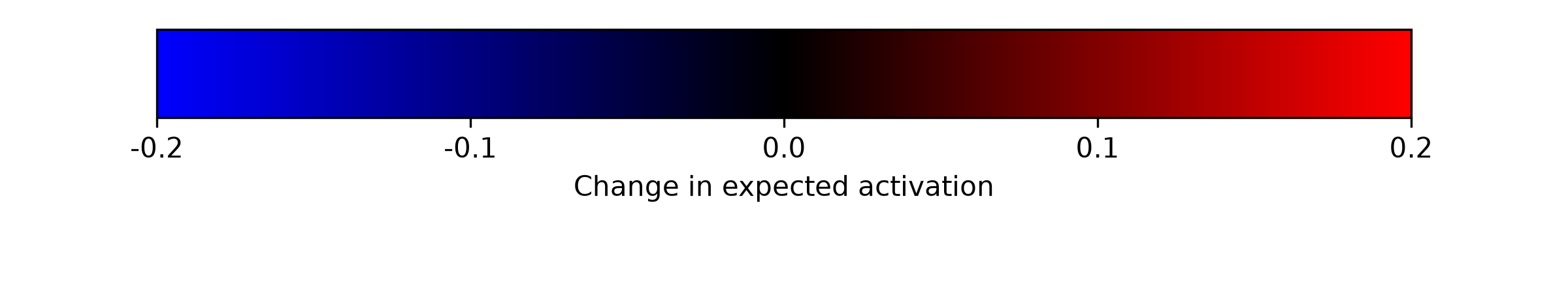}\par\smallskip
  \makebox[0.05\linewidth]{(b)}\includegraphics[width=0.78\linewidth]{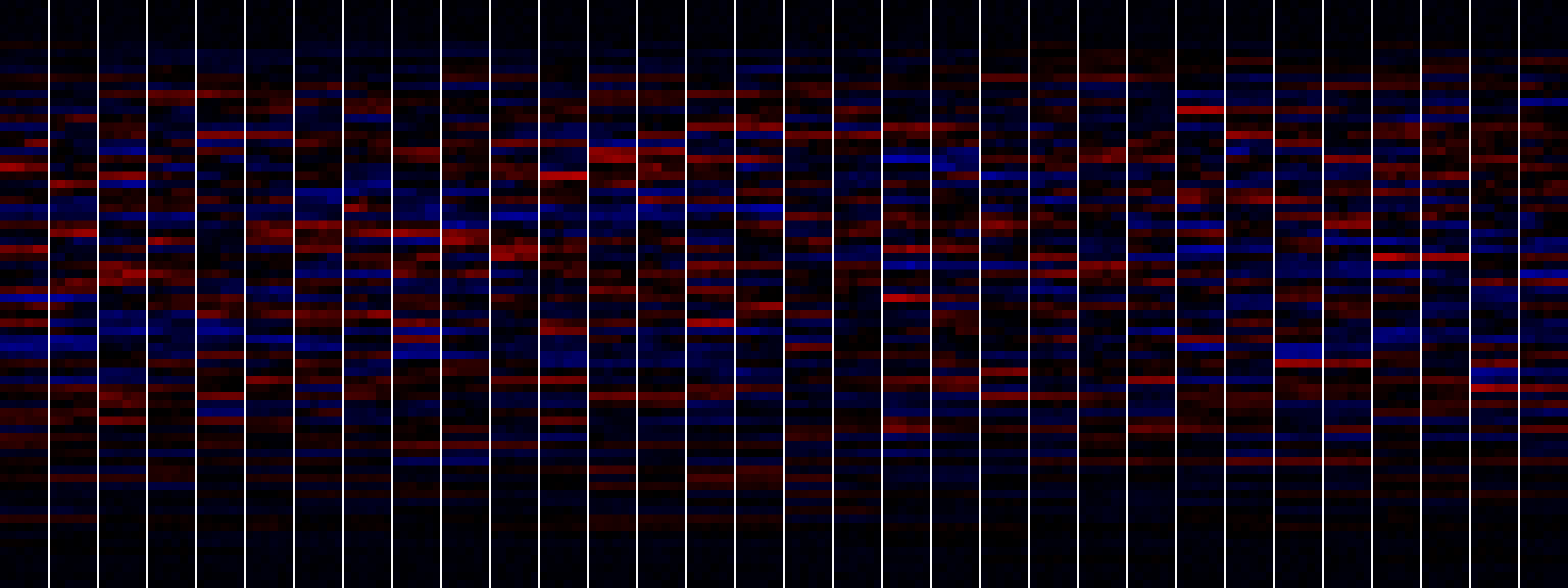}\par\smallskip
  \makebox[0.05\linewidth]{(c)}\includegraphics[width=0.78\linewidth]{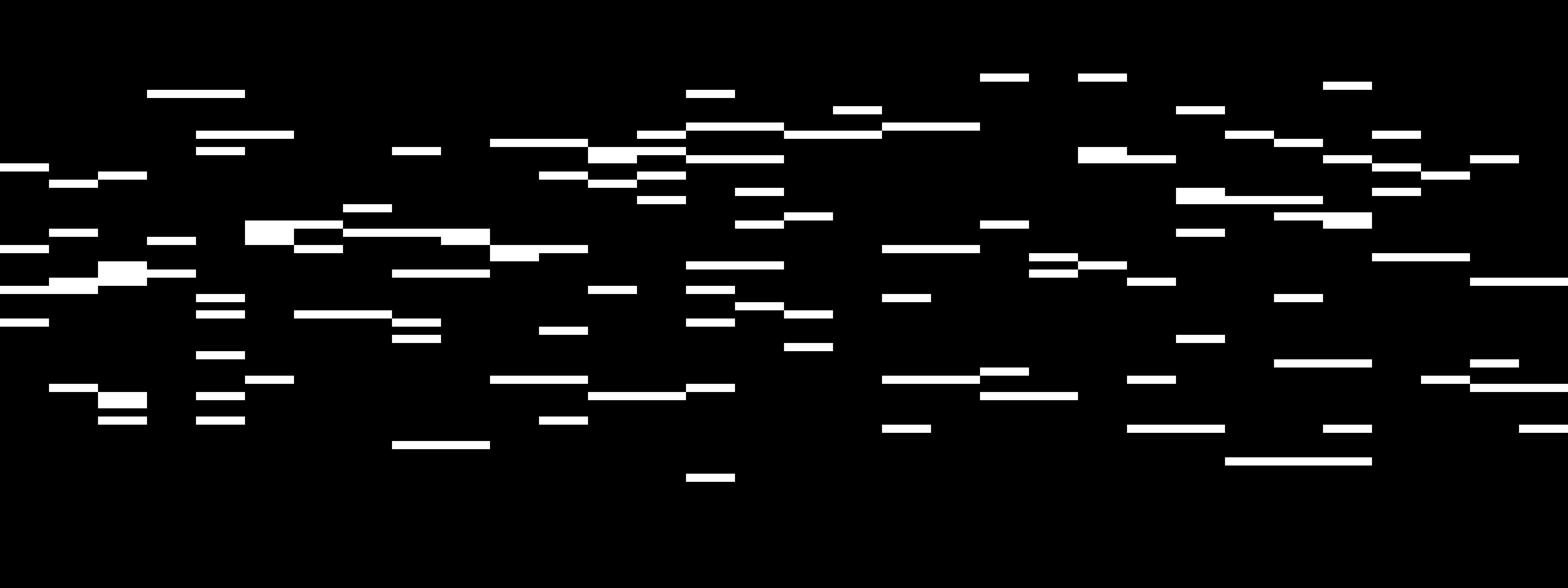}
  \caption{(Color online) Change in visible-layer expectations induced by activating the first hidden unit, relative to the all-inactive hidden configuration. (a) Difference map \(\sigma(b_i+w_{i1})-\sigma(b_i)\), with positive changes in red, negative changes in blue, and zero in black. The color scale ranges from \(-0.2\) to \(0.2\). (b) The same map with vertical lines at sixteenth-note intervals. (c) Binary representation obtained by averaging the differences over each sixteenth-note interval within each pitch row and displaying means of \(0.05\) or greater in white and all others in black.}
  \label{fig:one_hot}
\end{figure}

To investigate what kinds of patterns are associated with individual hidden units, we compared the visible expectations obtained with one hidden unit active and all others inactive against those obtained with all hidden units inactive. For hidden unit \(j\), the displayed difference is \(\Delta_j\langle v_i\rangle = \sigma(b_i+w_{ij})-\sigma(b_i)\). This subtracts the visible-bias baseline and isolates the change induced by activating the selected hidden unit, while retaining the trained parameters. Figure~\ref{fig:one_hot} shows the first hidden unit as a representative example. Positive differences are shown in red, negative differences in blue, and zero in black, on a fixed scale from \(-0.2\) to \(0.2\). Local temporal patterns with a width comparable to a sixteenth-note duration are visible in this difference map. Vertical lines mark these intervals, and a binary panel highlights pitch--time groups whose mean difference over a sixteenth-note interval is at least \(0.05\).

Although Fig.~\ref{fig:one_hot} shows only one representative unit, we performed the same difference-map analysis for all hidden units. Across these visualizations, pitch patterns with a duration comparable to a sixteenth note were observed, but recognizable chordal structures were not identified. Typical melodic phrases were also scarcely observed. In contrast, chordal structures were observed in the generated music, leaving open the possibility that chordal information is encoded through the coordinated activity of multiple hidden units. Such distributed representations are compatible with the mixtures of features discussed in previous work~\citep{Fernandez2023}.

Our single-unit analysis was motivated by the stroke-like features associated with individual hidden units in MNIST-trained RBMs~\citep{Liao2022}. By analogy, we expected recognizable musical components, such as chords or characteristic phrases, to be visible in individual-unit responses. The present analysis did not reveal such clear musical counterparts of strokes, despite the local pitch--time patterns observed. This result limits the single-unit analogy and does not establish an absence of chordal information from the hidden layer as a whole.

\subsection{Hidden-layer Representations under Transposition}

To examine how the trained RBM internally represents musical structure, we analyzed its responses to transposed musical inputs. In an RBM, information presented to the visible layer is compressed and encoded in the hidden layer, from which the visible states can be reconstructed. The hidden layer was therefore examined to characterize internal representations of musical inputs.

Among the 56 compositions used for training, two pieces (BWV853 and BWV865) were selected, and versions shifted by a semitone and a whole tone were created within the same transposition range used for training. For each version, piano-roll images were generated and segmented into multiple two-measure vectors, which were used as inputs to the visible layer. Corresponding hidden-unit activations were sampled and analyzed.

\begin{figure}[htbp]
  \centering
  \begin{picture}(0,0)
    \put(-10,200){\small (a)}
  \end{picture}
  \includegraphics[width=0.48\linewidth]{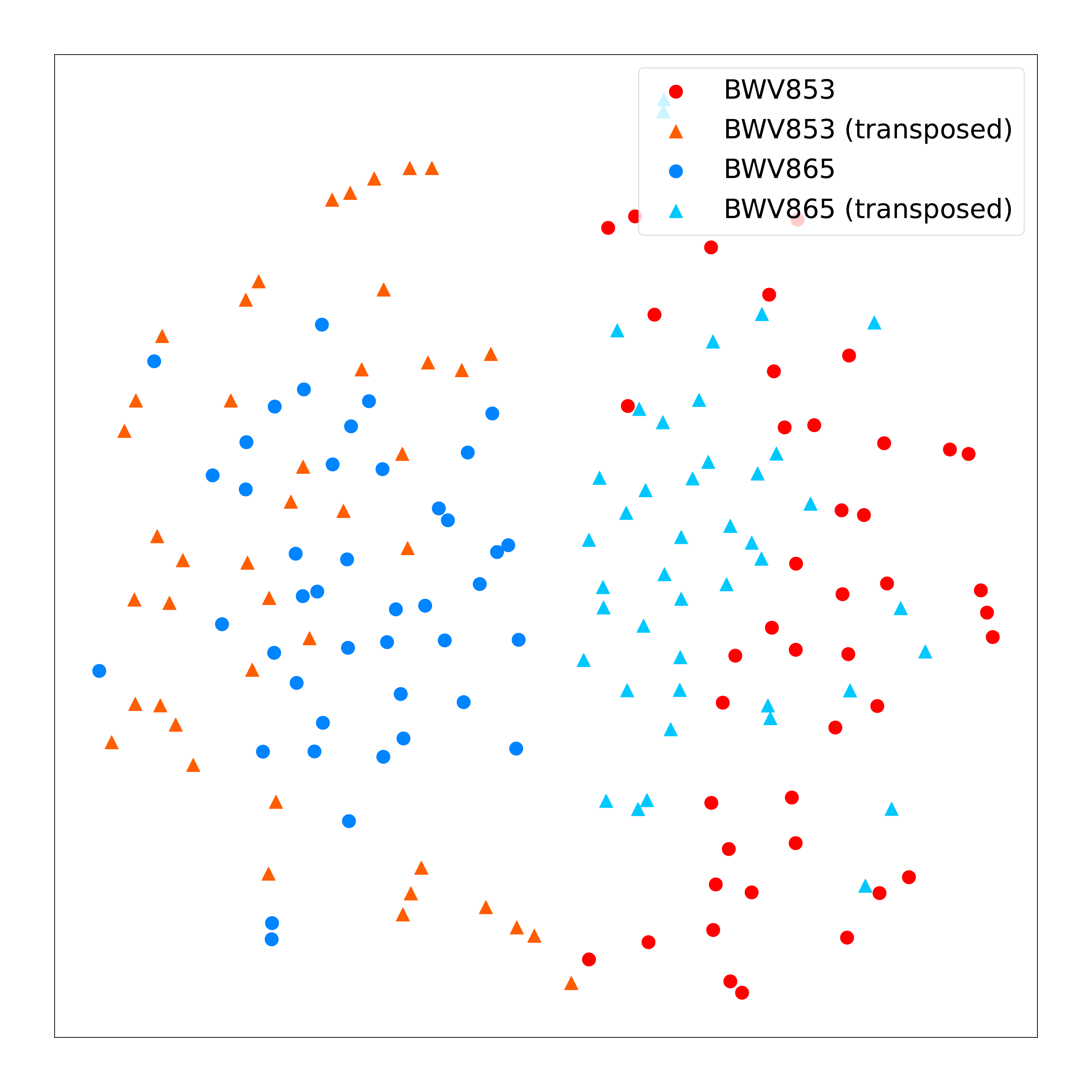}
  \begin{picture}(0,0)
    \put(-10,200){\small (b)}
  \end{picture}
  \includegraphics[width=0.48\linewidth]{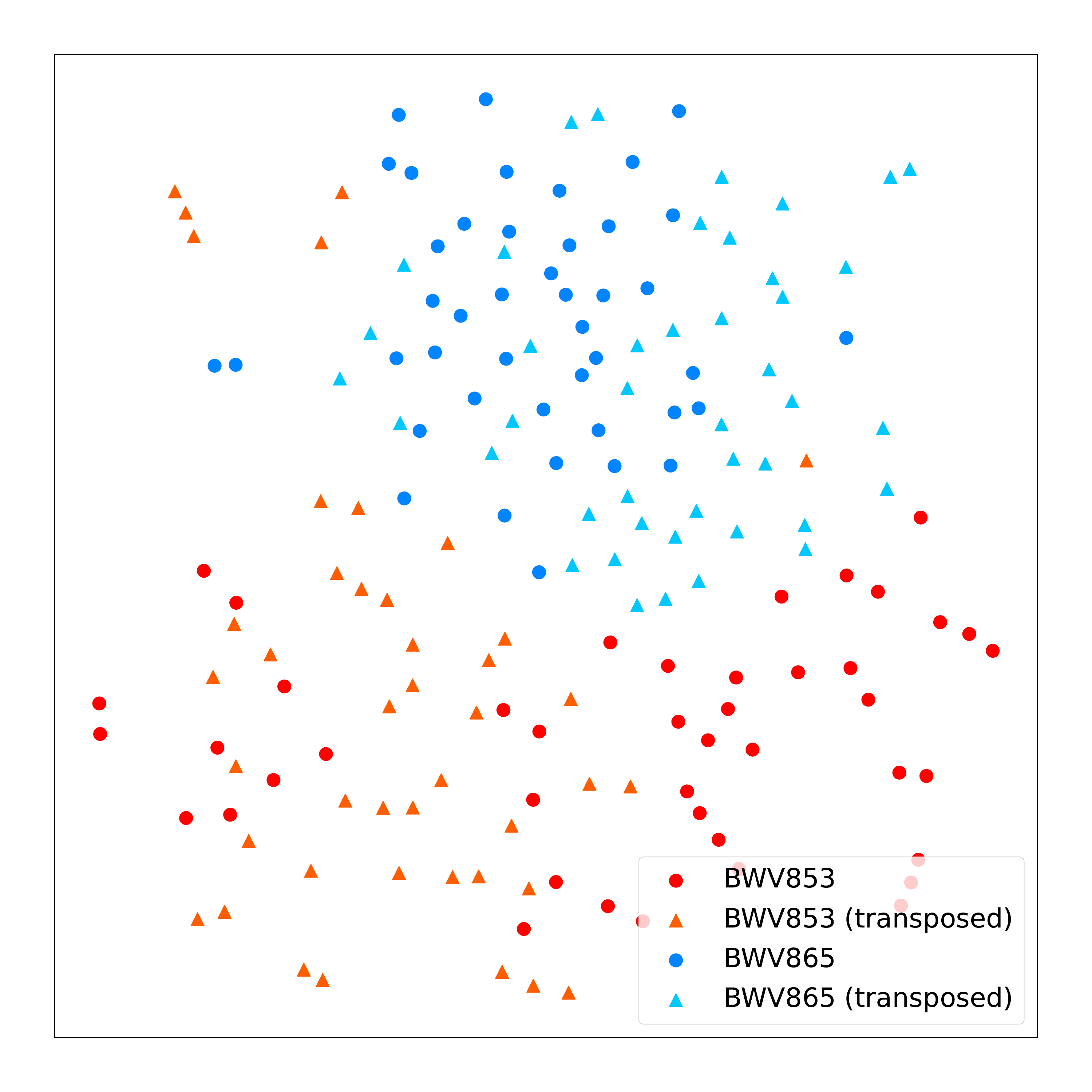}
  \caption{(Color online) The hidden layer representations were projected into two dimensions using t-SNE. Original BWV~853 and BWV~865 inputs are shown as red and blue circles, respectively, and their transposed versions as orange and light-blue triangles. (a) Transposition by a semitone. (b) Transposition by a whole tone.}
  \label{fig:hidden}
\end{figure}

Figure~\ref{fig:hidden} shows the results of dimensionality reduction of the hidden activations using t-SNE~\citep{Hinton2008}. In the semitone-transposition visualization, the original and transposed inputs occupy different regions (Fig.~\ref{fig:hidden}(a)), whereas whole-tone transposition produces greater proximity between original and transposed inputs (Fig.~\ref{fig:hidden}(b)). These visual impressions require quantitative examination in the original hidden space, since proximity in the two-dimensional projection need not preserve all high-dimensional similarities.

To quantify and discuss the t-SNE results, we performed a cosine-similarity analysis in both the input and original hidden spaces. For vectors \(\mathbf{x}\) and \(\mathbf{y}\), we use \(s(\mathbf{x},\mathbf{y})=\mathbf{x}\cdot\mathbf{y}/(\|\mathbf{x}\|\|\mathbf{y}\|)\). Each group contains 43 two-measure windows, and each group comparison includes all \(43\times43=1849\) pairs, including corresponding time intervals. The input vectors are flattened binary piano rolls, and the hidden vectors are the sampled binary states used for Fig.~\ref{fig:hidden}, before t-SNE projection. These hidden states were obtained using the reference implementation's 0/255 input convention; uniformly scaling the input vectors from 0/1 to 0/255 does not change their input-space cosine similarities. Table~\ref{tbl:cosine_transposition} reports the mean and sample standard deviation across pairs. The pairs share windows and are not independent, so the standard deviations describe the spread of similarities rather than uncertainty in the means.

For semitone transposition, the cross-piece combinations of an original and a transposed input have the highest mean similarities in both spaces. For whole-tone transposition, the original and transposed versions of the same piece have comparatively high similarities in both spaces. The correspondence is qualitative rather than an identical ranking of all pairs. In particular, the semitone grouping suggested visually by t-SNE is not fully reproduced by the mean cosine similarities in the original hidden space.

\begin{table}[htbp]
  \centering
  \small
  \setlength{\tabcolsep}{4pt}
  \caption{Cosine similarities between input and hidden groups under transposition. A and B denote BWV~853 and BWV~865, respectively; primes denote their transposed versions. Each entry is the mean $\pm$ sample standard deviation across 1,849 cross-group window pairs. Hidden vectors are the binary samples used for Fig.~\ref{fig:hidden}.}
  \label{tbl:cosine_transposition}
  \begin{tabular}{lrrrr}
    \hline
           & \multicolumn{2}{c}{Semitone} & \multicolumn{2}{c}{Whole tone}                                             \\
    Pair   & Input                        & Hidden                         & Input               & Hidden              \\
    \hline
    A--B   & $0.0459 \pm 0.0314$          & $0.2197 \pm 0.0258$            & $0.0459 \pm 0.0314$ & $0.2197 \pm 0.0258$ \\
    A'--B' & $0.0459 \pm 0.0314$          & $0.2195 \pm 0.0260$            & $0.0459 \pm 0.0314$ & $0.2196 \pm 0.0261$ \\
    A--A'  & $0.0489 \pm 0.0396$          & $0.2309 \pm 0.0265$            & $0.0861 \pm 0.0545$ & $0.2429 \pm 0.0339$ \\
    B--B'  & $0.0587 \pm 0.0360$          & $0.2151 \pm 0.0281$            & $0.0969 \pm 0.0517$ & $0.2279 \pm 0.0354$ \\
    A--B'  & $0.1070 \pm 0.0516$          & $0.2454 \pm 0.0340$            & $0.0665 \pm 0.0406$ & $0.2266 \pm 0.0280$ \\
    A'--B  & $0.1048 \pm 0.0508$          & $0.2448 \pm 0.0341$            & $0.0655 \pm 0.0416$ & $0.2257 \pm 0.0292$ \\
    \hline
  \end{tabular}
\end{table}

As a simpler comparison based on second-order statistics, we applied principal component analysis (PCA) to the same training piano rolls used for the RBM. The 21,444 binary training windows were centered by subtracting the mean of each pixel, without scaling by the pixel standard deviation or whitening. We then projected the same 43 windows per group onto the first two principal components. A single PCA basis and common axis ranges were used for both transposition conditions. PC1 and PC2 explain approximately 1.75\% and 1.75\% of the training variance, respectively, or 3.49\% in total.

\begin{figure}[htbp]
  \centering
  \begin{minipage}[t]{0.49\linewidth}
    \raggedright (a)\par
    \centering\includegraphics[width=\linewidth]{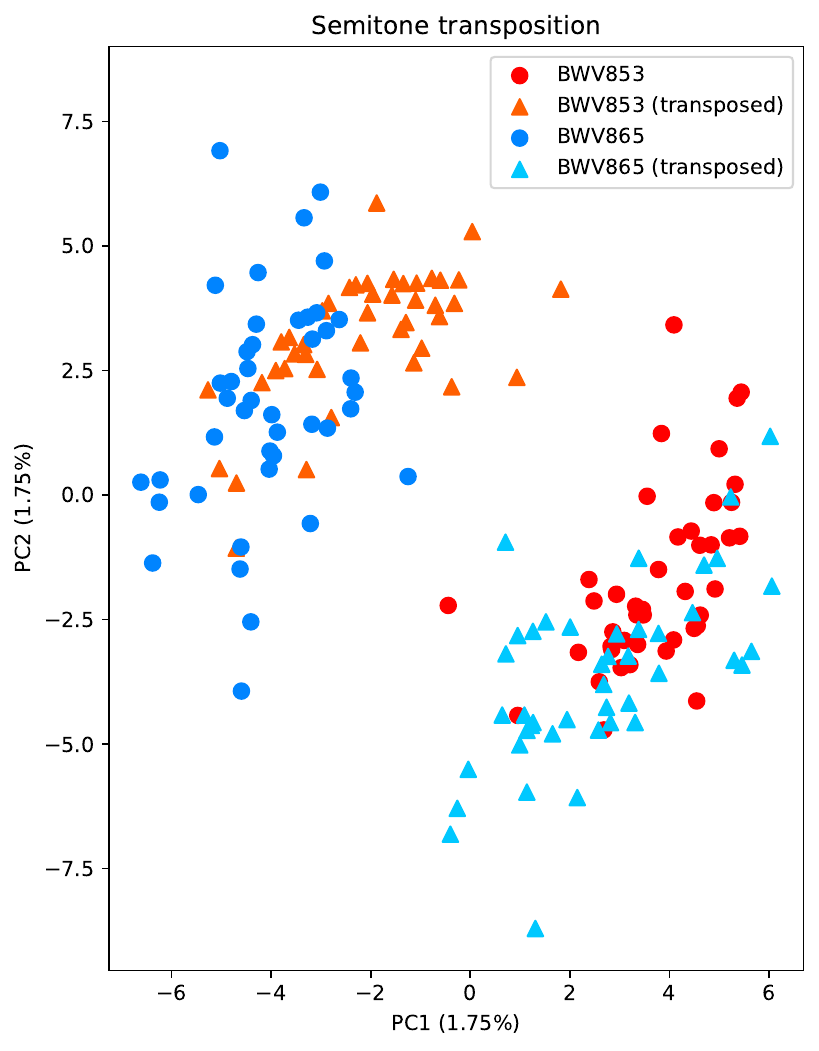}
  \end{minipage}\hfill
  \begin{minipage}[t]{0.49\linewidth}
    \raggedright (b)\par
    \centering\includegraphics[width=\linewidth]{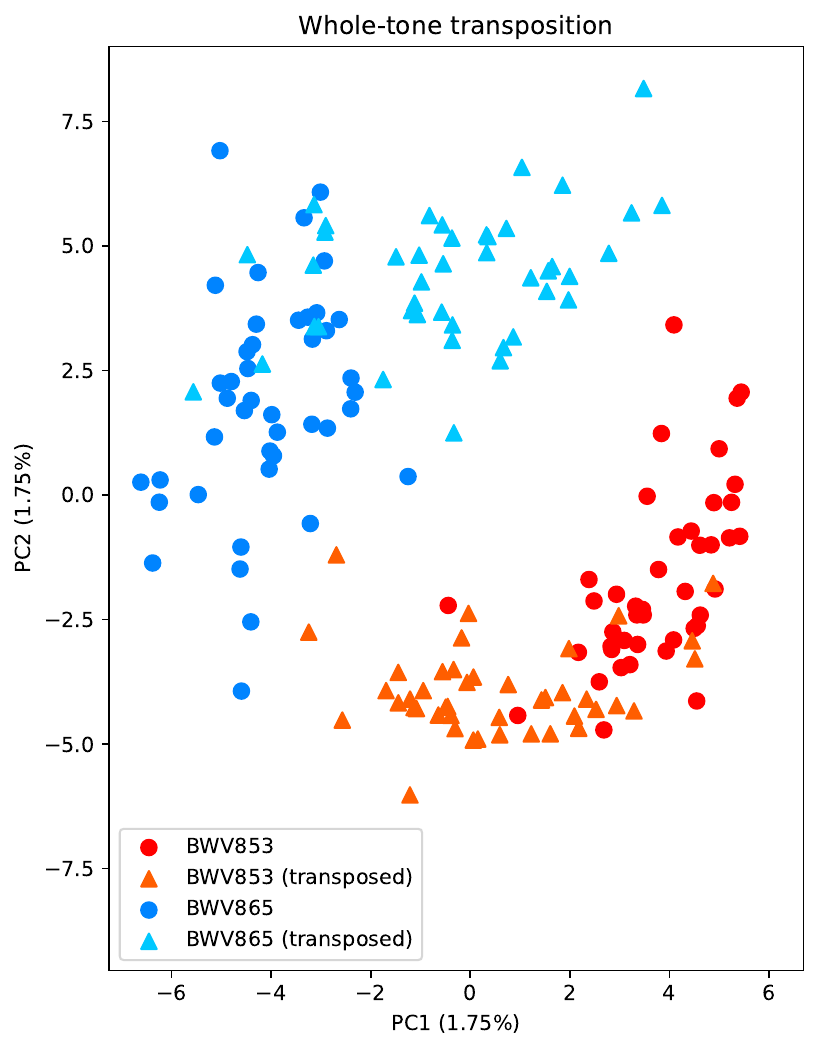}
  \end{minipage}
  \caption{(Color online) PCA projections of the input piano rolls used in Fig.~\ref{fig:hidden}. Original BWV~853 and BWV~865 inputs are shown as red and blue circles, respectively, and their transposed versions as orange and light-blue triangles. (a) Semitone transposition. (b) Whole-tone transposition. Both panels use the same PCA basis fitted to the full RBM training dataset and the same axis ranges. PC1 and PC2 together explain approximately 3.49\% of the training variance.}
  \label{fig:pca_transposition}
\end{figure}

Figure~\ref{fig:pca_transposition} shows that transposition-dependent groupings also arise directly from the input piano rolls under PCA. In the semitone condition, the original inputs of one piece overlap with transposed inputs of the other piece, whereas whole-tone transposition gives greater proximity between original and transposed inputs of the same piece. Thus, similar qualitative organization can be obtained without using an RBM hidden representation.

Taken together, the shared tendencies in input and hidden cosine similarities and the PCA comparison suggest that the organization seen in t-SNE reflects simple overlap between input patterns and their second-order statistics. The observed grouping therefore does not by itself demonstrate recognition of transposition-equivalent musical phrases beyond these input properties. This interpretation is consistent with sensitivity to absolute pitch positions and with the lack of built-in translation invariance in the standard RBM architecture~\citep{Lee2009}. However, the two displayed principal components account for only a small fraction of the total variance, and this comparison neither establishes a complete explanation of the hidden representation nor excludes relative-pitch information in other components or distributed hidden states.

\subsection{Samples from the trained RBM}

\begin{figure}[htbp]
  \centering
  \begin{tabular}{@{}l@{\hspace{1em}}l@{\hspace{1em}}l@{}}
    {\small (a)} & {\small (b)} & {\small (c)} \\[2pt]
    \includegraphics[width=0.27\linewidth]{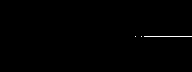} &
    \includegraphics[width=0.27\linewidth]{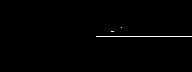} &
    \includegraphics[width=0.27\linewidth]{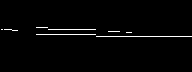} \\[6pt]
    {\small (d)} & {\small (e)} & {\small (f)} \\[2pt]
    \includegraphics[width=0.27\linewidth]{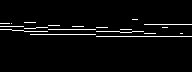} &
    \includegraphics[width=0.27\linewidth]{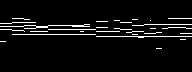} &
    \includegraphics[width=0.27\linewidth]{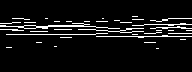}
  \end{tabular}
  \caption{Progression of the visible configuration over the generation steps of Algorithm~\ref{proc:algorithm1}. Images (a) through (f) contain \( t = 50, 100, 250, 500, 750, \) and \( 1000 \) active visible units, respectively.}
  \label{fig:music_composition}
\end{figure}

\begin{figure}[htbp]
  \centering
  \begin{picture}(12,8)
    \put(0,135){\small (a)}
  \end{picture}
  \includegraphics[width=0.95\linewidth]{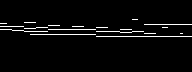}
  \begin{picture}(12,8)
    \put(0,45){\small (b)}
  \end{picture}
  \includegraphics[width=0.95\linewidth]{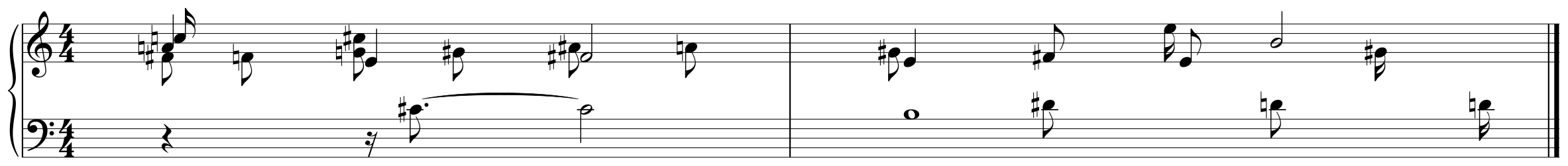}
  \caption{A two-measure piano-roll configuration with 496 active visible units, obtained at the energy minimum along the generation trajectory. (a) Piano roll representation of the sample. (b) Sheet music representation of the same sample, with note onsets and offsets quantized to a sixteenth-note grid.}
  \label{fig:pianoroll_minimum}
\end{figure}

\begin{figure}[htbp]
  \centering
  \includegraphics[width=10cm]{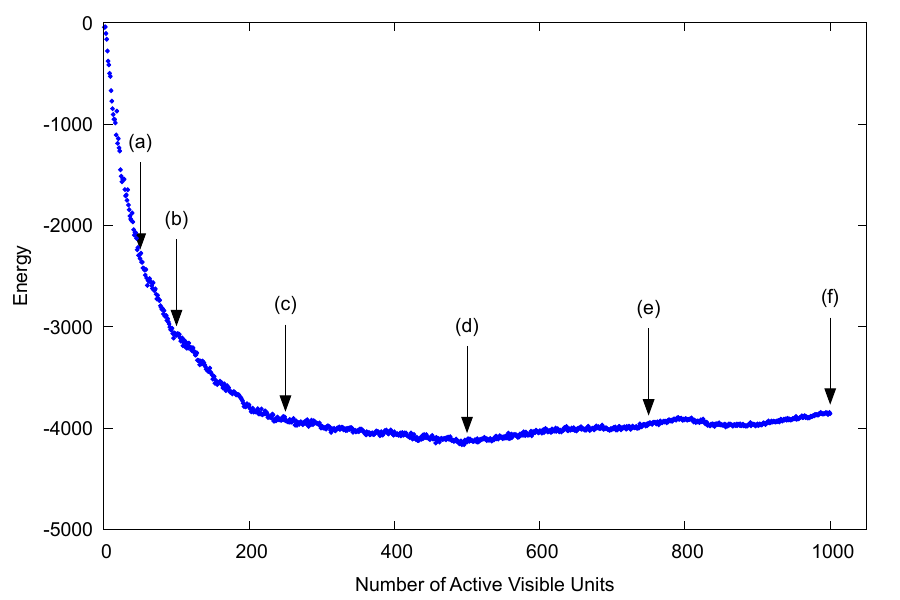}
  \caption{RBM energy as a function of the number of active visible units in the configurations generated by Algorithm~\ref{proc:algorithm1}. The number of active visible units is controlled by selecting the largest conditional expectations, rather than by a Gibbs draw of the visible state. Labels (a)-(f) correspond to images (a)-(f) shown in Fig.~\ref{fig:music_composition}.}
  \label{fig:energy_music}
\end{figure}

An example of two-measure sampling using Algorithm~\ref{proc:algorithm1} is shown in Fig.~\ref{fig:music_composition}. The figure shows the visible states \(\mathbf{v}_t\) at sampling steps \( t = 50, 100, 250, 500, 750, \) and \( 1000 \). All images exhibit the structure of piano rolls. The time evolution of the RBM energy during this sampling procedure is shown in Fig.~\ref{fig:energy_music}.
The energy decreases up to approximately 500 active visible units, after which it begins to increase. Along this generation trajectory, configurations with either too few or too many active pixels (notes) have higher energies. We refer to the activity level at the energy minimum along the trajectory as the optimal number of active units for this generation algorithm. The energy reached its minimum at \( t = 496 \) active visible units. The corresponding piano roll is shown in Fig.~\ref{fig:pianoroll_minimum}. For the sheet music in Fig.~\ref{fig:pianoroll_minimum}(b), note onsets and offsets were quantized to the nearest sixteenth-note grid, with a minimum duration of one sixteenth note. The sample contains local simultaneous pitch combinations corresponding to C\# diminished and minor triads, F\# major and minor triads, and E major and B major triads. In particular, E--G\#--B and B--D\#--F\# occur successively in the second measure. These observations indicate local chordal organization, but do not establish a single key or a coherent functional harmonic progression across the entire sample.

The training piano rolls contain an average of \(246.6 \pm 0.4\) active pixels per measure, corresponding to approximately 493, or about 500, active pixels per two-measure window. The quoted uncertainty is the standard error of the mean calculated by treating the measure entries in the training windows as independent; it does not account for correlations due to overlapping windows and transposition augmentation. The low-energy region near 500 active visible units is therefore consistent with the average activity in the training data, although this agreement alone does not determine the energy-minimizing activity level.

Different random seeds during generation produced markedly different musical sequences, but the energy dependence on the number of active visible units remained nearly unchanged. The optimal number also remained similar when the RBM was retrained with different random seeds. Thus, although this optimum is specific to the generation algorithm and does not establish the typical activity of the RBM distribution, it appears reasonably robust across the seeds examined when the number of hidden units and the training dataset are held fixed.

\begin{figure}[htbp]
  \centering
  \begin{picture}(12,8)
    \put(0,28){\small (a)}
  \end{picture}
  \includegraphics[width=0.95\linewidth]{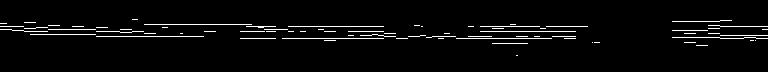}
  \begin{picture}(12,8)
    \put(0,110){\small (b)}
  \end{picture}
  \includegraphics[width=0.95\linewidth]{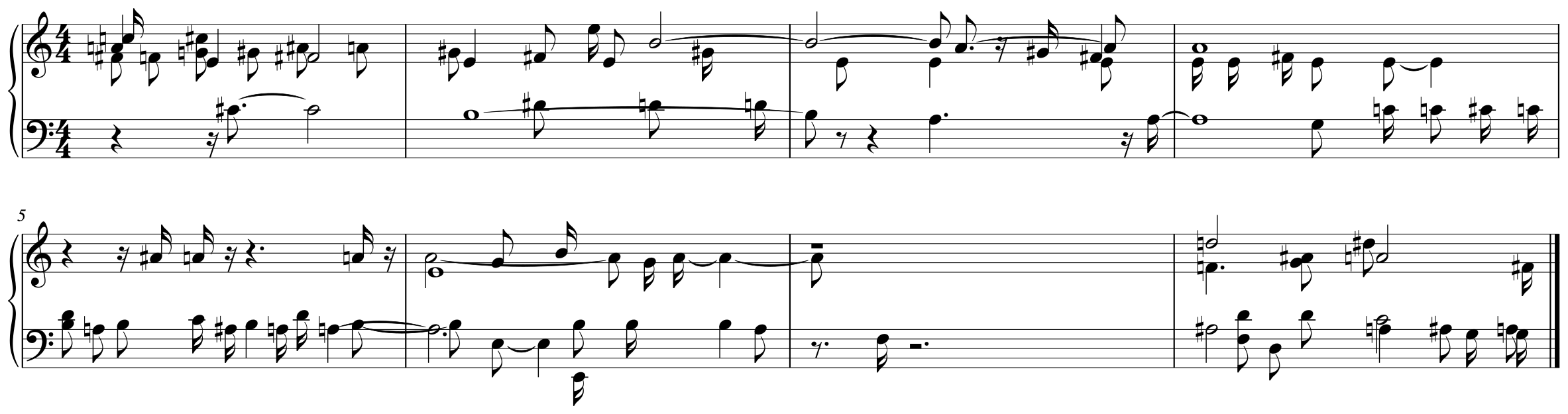}
  \caption{An 8-measure piano-roll configuration obtained by iterative continuation sampling from the trained RBM. (a) Piano roll representation of the 8-measure sample. (b) Sheet music representation of the same sample, with note onsets and offsets quantized to a sixteenth-note grid.}
  \label{fig:chain_composition}
\end{figure}

An example of an eight-measure piano roll obtained by the continuation procedure in Algorithm~\ref{proc:algorithm2} is shown in Fig.~\ref{fig:chain_composition}. This image also exhibited a piano roll structure, similar to the two-measure images shown in Fig.~\ref{fig:pianoroll_minimum}. Local pitch patterns remain visible in this example, but they do not establish long-range musical coherence under the present continuation algorithm. As noted in the algorithm description, the loss of coherence is likely to depend on the generation procedure and cannot be attributed to the RBM alone. Audio examples from earlier generation experiments are available online~\citep{rbm-music-demo}.

\section{Summary and Discussion}\label{sec:summary}

In this study, we used piano-roll representations of symbolic music as structured binary data and analyzed the internal representations learned by a Bernoulli-Bernoulli RBM. The trained RBM reconstructed piano-roll configurations, including examples derived from musical pieces not used during training, and assigned lower visible free energies to the examined piano-roll configurations than to the digit and noise inputs. These free-energy differences establish higher probabilities for these particular piano-roll inputs under the trained model, but do not distinguish learning musical relationships from learning sparsity or horizontal continuity, nor demonstrate general-purpose compositional ability.

The simplicity of the RBM architecture allowed us to analyze the learned representation directly. By fixing individual hidden units and computing the corresponding visible-layer expectations, we found that hidden units induce sparse piano-roll-like patterns and local temporal or pitch-statistical structures. However, these patterns were not cleanly separated into musical concepts such as melodies, chords, or keys. The t-SNE analysis, input--hidden cosine-similarity comparison, and PCA of the input piano rolls suggest that the observed organization reflects simple overlap between input patterns and their second-order statistics, with sensitivity to absolute pitch positions. These results do not by themselves establish whether relative-pitch relationships are absent from the representation. This behavior is consistent with previous observations that RBMs and Deep Belief Networks lack inherent translational invariance in their input space. In contrast, convolutional deep belief models, which incorporate local receptive fields and weight sharing, offer a possible route toward improved recognition of translated or transposed patterns~\citep{Lee2009}.

The sampled piano-roll configurations should therefore be interpreted as probes of the learned energy landscape, not as evidence for a competitive music generation method. Short samples showed piano-roll-like sparsity and some local pitch organization, whereas long-range musical coherence was not reliably maintained under the present continuation algorithm. This loss of coherence may depend substantially on the generation rule and cannot be attributed to the RBM alone. The hidden-unit and transposition analyses suggest that the examined models capture local pitch-time statistics and absolute-pitch regularities, while higher-level musical structures were not clearly separated into individual hidden units by the present analysis. Future work may examine whether these learned structures are reflected in quantitative properties of the weight matrix, such as singular value spectra, as suggested in recent studies of RBM tasks~\citep{Ichikawa2022}. Moreover, previous work has indicated that hidden units in RBMs can encode prototypical patterns in the visible layer~\citep{Hinton2002}, although such patterns are often difficult to interpret in standard RBMs. Architectures such as classification RBMs, which tend to produce more distinguishable hidden-unit activations~\citep{Larochelle2012}, may facilitate the identification of prototypical melodic or harmonic structures, and exploring such architectural extensions remains an important topic for future investigation.

Training on the same dataset with different random seeds and varying the generation seed yielded broadly similar qualitative behavior, although the generated musical sequences differed. The optimal activity level under our generation algorithm remained approximately unchanged, suggesting a connection to the activity of the training data. Varying the number of hidden units also produced similar qualitative behavior once sufficient capacity was available; models with too few hidden units failed to adequately reconstruct the piano rolls. Thus, the observed behavior did not depend strongly on the precise hidden-layer size within the sufficient-capacity range examined. These checks do not exclude effects of dataset size, training effectiveness, or the analysis and generation procedures, and our conclusions are limited to the experimental conditions examined here.


\let\doi\relax


\bibliographystyle{ptephy}
\bibliography{reference}

\appendix

\section{Piano-Roll Generation by Standard Gibbs Sampling}
\label{sec:standard_gibbs}

To examine the behavior of the trained RBM under ordinary block Gibbs sampling, we used two initial visible configurations. One was an image whose pixels were independently set to 0 or 1 with equal probability, and the other was an all-zero image. At each Gibbs step, we sampled the hidden units conditionally on the current visible state and then sampled the visible units conditionally on the hidden state. No constraint was imposed on the number of active visible units, and no selection of pixels by their conditional expectations was performed. Each chain was run for 1,000 Gibbs steps. The joint energy was evaluated after each visible update using the hidden state that generated it; step 0 denotes the initial visible configuration with a conditionally sampled hidden state.

\begin{figure}[htbp]
  \centering
  \begin{minipage}[t]{0.85\linewidth}
    \raggedright (a)\par
    \centering\includegraphics[width=\linewidth]{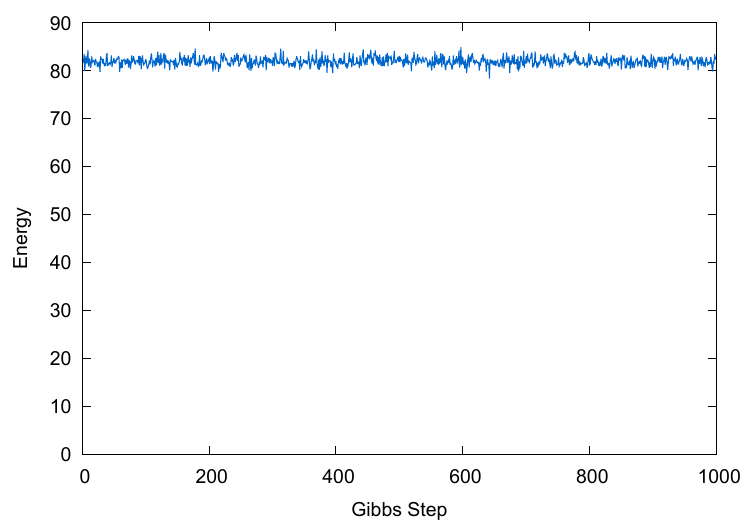}
  \end{minipage}
  \par\smallskip
  \begin{minipage}[t]{0.85\linewidth}
    \raggedright (b)\par
    \centering\includegraphics[width=\linewidth]{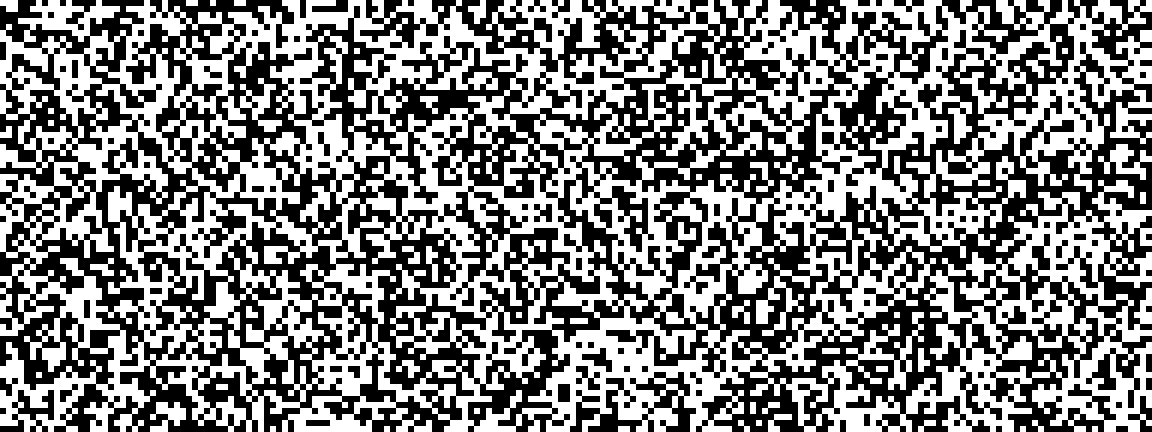}
  \end{minipage}
  \caption{Standard block Gibbs sampling initialized with independent random binary pixels. (a) Joint RBM energy as a function of the Gibbs step. (b) Visible configuration after 1,000 Gibbs steps.}
  \label{fig:gibbs_from_random}
\end{figure}

\begin{figure}[htbp]
  \centering
  \begin{minipage}[t]{0.85\linewidth}
    \raggedright (a)\par
    \centering\includegraphics[width=\linewidth]{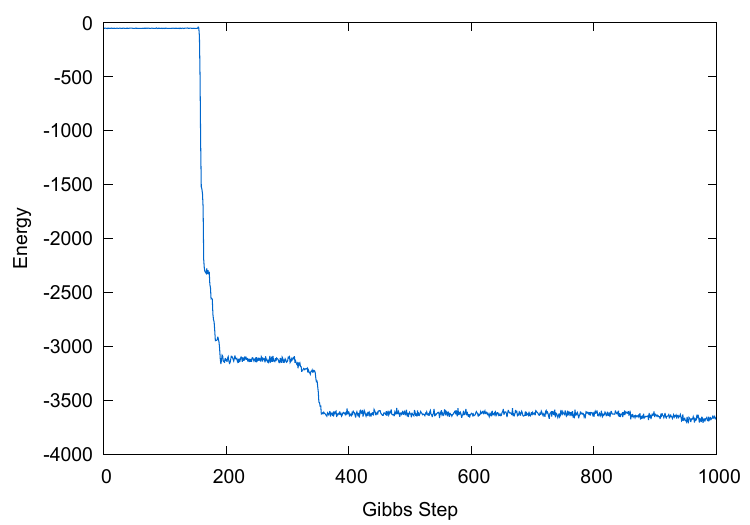}
  \end{minipage}
  \par\smallskip
  \begin{minipage}[t]{0.85\linewidth}
    \raggedright (b)\par
    \centering\includegraphics[width=\linewidth]{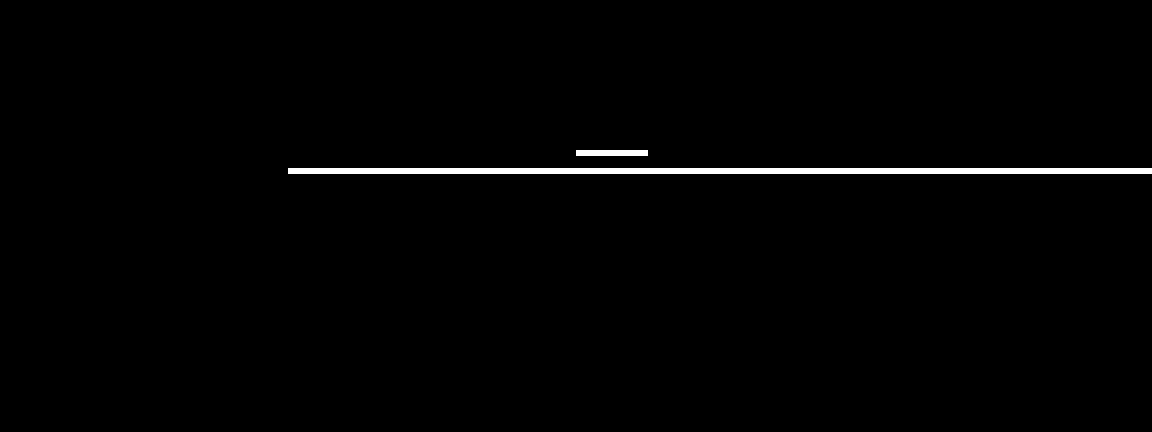}
  \end{minipage}
  \caption{Standard block Gibbs sampling initialized with an all-zero image. (a) Joint RBM energy as a function of the Gibbs step. (b) Visible configuration after 1,000 Gibbs steps.}
  \label{fig:gibbs_from_black}
\end{figure}

Figures~\ref{fig:gibbs_from_random} and \ref{fig:gibbs_from_black} show the energy traces and the final visible configurations. With random initialization, the output remained noise-like. With all-zero initialization, the energy decreased, but the final image contained only sparse horizontal traces. In these trials, standard Gibbs sampling did not produce a coherent musical configuration. These examples concern the specified initial conditions and finite sampling length; they do not establish that standard Gibbs sampling cannot generate musical structures under other conditions.

\clearpage
\section{Reconstruction of MNIST Images}
\label{sec:mnist_reconstruction}

\begin{figure}[htbp]
  \centering
  \begin{tabular}{@{}l@{\hspace{0.02\linewidth}}l@{}}
    \includegraphics[width=0.49\linewidth]{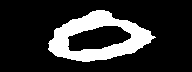} &
    \includegraphics[width=0.49\linewidth]{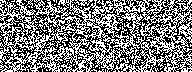} \\
    \includegraphics[width=0.49\linewidth]{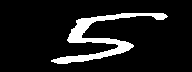} &
    \includegraphics[width=0.49\linewidth]{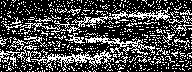} \\
    \includegraphics[width=0.49\linewidth]{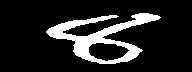} &
    \includegraphics[width=0.49\linewidth]{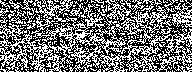}
  \end{tabular}
  \caption{Outputs obtained by Gibbs reconstruction from digit-image inputs using the trained RBM. Panels (a), (b), and (c) show input images from the MNIST dataset. Panels (a'), (b'), and (c') show the corresponding output images obtained by Gibbs reconstruction. The outputs are noise-like rather than digit-like, consistent with the trained energy model being specific to piano-roll-like visible configurations.
  }
  \label{fig:reconstruct_MNIST}
\end{figure}

To evaluate whether the RBM trained on piano rolls can reconstruct images outside the training domain, we used the MNIST dataset as input. Each \( 28\times 28\) pixel image was resized to \(72 \times 192\) pixels and provided to the visible units. As in the piano-roll reconstruction, the visible state is initialized with the binarized input image, followed by one conditional Bernoulli draw of the hidden state and one conditional Bernoulli draw of the visible state. The latter visible sample is the reconstruction, so exactly one block Gibbs step is performed with the trained parameters fixed, without thresholding the output expectations or constraining the number of active pixels. The result is shown in Fig.~\ref{fig:reconstruct_MNIST}. In contrast to the case of piano roll images in Fig.~\ref{fig:reconstruct_piano_roll}, the RBM failed to reconstruct digit images and instead produced noise-like outputs. These results indicate that the RBM trained on piano rolls can reconstruct unseen piano roll images more accurately than images that differ in nature, such as handwritten digits.

\end{document}